\documentclass[12pt,a4paper]{article}
\usepackage{setspace}
\usepackage{subfig}
\usepackage{authblk}
\usepackage{amssymb}

\usepackage{listings}
\usepackage{caption}
\usepackage{amsmath, epsfig,graphicx}
\usepackage{color}
\usepackage{enumerate}

\def\Title#1{\begin{center} {\Large \bf #1 } \end{center}}
\def\Abstracti#1{\begin{center} {\small \bf #1 } \end{center}}
\def\Author#1{\begin{center}{ \sc #1} \end{center}}
\def\Address#1{\begin{center}{ \it #1} \end{center}}
\newenvironment{Abstract}{\begin{quotation}  }{\end{quotation}}

\begin{document}

\begin{titlepage}
	\let\thefootnote\relax\footnote{* Mansouri@quantgates.co.uk}
	
	\Title{}
	\Title{}
	\Title{Supersymmetry with Cadabra}
	\vfil
	\Author{ Behzad Mansouri*}
	\vfil
	\Address{QuantGates Ltd, 
		London, EC2A 4NE, UK}
	\vfil
	\Abstracti {Abstract}
	\begin{Abstract}
		This article provides a quick guide to the implementation of supersymmetry with Cadabra, a symbolic computer algebra system. Details are provided on the implementation of Grassmann variables, fermionic fields, supercharges, and superderivatives and the treatment of superfield expressions and supersymmetric Lagrangians. Finally, the automation of supersymmetric Lagrangian generation in the Cadabra programming framework is discussed.
		
	\end{Abstract}
	\vfill

\end{titlepage}

\section{Introduction to Supersymmetry and Cadabra}
Supersymmetry, strings and branes are conjectured to be key ingredients toward establishing a unique unified theory of physics. In addition to the extension of possible symmetry groups relating bosons to fermions, new structures within the theory seem to provide approaches to treating some conceptual problems in high-energy physics, including the hierarchy problem.

The underlying calculations in supersymmetric theories, especially in higher dimensions, are technically long, and using a symbolic computer algebra system for the implementation of such highly structured theories can facilitate calculations in the model under study. Cadabra is a new computer algebra system (CAS) that was developed specifically for solutions to problems that have emerged in field theory. It was first developed to study the supersymmetry of higher-derivative effective actions \cite{1,2}. The extensive capabilities of Cadabra, such as the ability to deal with any type of tensor, anticommuting variables, Clifford algebras and Fierz transformations, as well as its many ready-to-run simplification algorithms and fully programmable features, make it a powerful tool to efficiently study problems emerging in quantum field theory and string theory. In \cite{3,4,5}, Cadabra was introduced, and the various capabilities of this new computer algebra system, such as its comprehensive functionality for large-scale tensor computations, were discussed. In \cite{6,7,8,9,10},  Cadabra was extensively used to verify or derive results in the supergravity context. 

The aim of this paper is to illustrate how Cadabra is used for supersymmetric computations and provide the source code for a variety of explicit calculations, paving the way for better and faster routes to study supersymmetric aspects of string theory and M-theory in more depth via Cadabra.  

\subsection{Indices, Properties and Algorithms in Cadabra}

In Cadabra, all objects in the form of subscripts or superscripts are considered to be $indices$. Cadabra recognizes free and dummy indices, and it has an automatic index renaming feature. Cadabra uses LaTeX syntax.

\begin{verbatim}
	ex1:= \Gamma^{i j r} \Gamma^{k l m n} \Gamma_{r};
\end{verbatim}

We can assign mathematical properties to symbols:
\begin{verbatim}
\Gamma_{#}::GammaMatrix.
\partial{#}::PartialDerivative.
\end{verbatim}

With the above declaration, the properties of  \verb|GammaMatrix| and \verb|PartialDerivative|
are attached to the symbols $\Gamma$ and $\partial$, respectively.

After writing the mathematical expressions, computations can be executed on them by applying Cadabra's algorithms. 
There are many built-in algorithms for substitution and variation, index manipulations, sorting and simplifications, as well as those that have been designed to act on spinors and much more. In the following sections, we demonstrate the use of a variety of Cadabra's algorithms to perform different calculations.

\subsection{Gamma Matrices and Fierz transformations}
In this section, we show how to derive useful gamma matrices and Fierz identities within Cadabra. We define the indices and denote the underlying dimension and attach gamma matrix property to the $\Gamma$ symbol, and we also associate a metric with it:
\begin{verbatim}
{i,j,k,l,m,n}::Indices(vector).
{i,j,k,l,m,n}::Integer(0..7).
\Gamma_{#}::GammaMatrix(metric=\delta).
\delta_{m n}::KroneckerDelta.
\end{verbatim}

In Cadabra, simplification steps are contained in the function \verb|post_process|, which is executed on every new input. Based on our needs, we update it to consider \verb|sort_product|, \verb|eliminate_kronecker| and \verb|canonicalise|:

\begin{verbatim}
def post_process(ex):
    sort_product(ex)
    eliminate_kronecker(ex)
    canonicalise(ex)
    collect_terms(ex)
\end{verbatim}

We aim to find an identity for the below expression: 
\begin{verbatim}
ex1:= \Gamma^{i j m} \Gamma^{k l} \Gamma_{m};
\end{verbatim}

\begin{equation}
	\begin{array}{l}
{}\Gamma^{i j m} \Gamma^{k l} \Gamma_{m}
	\end{array}
\end{equation}

By executing \verb|join_gamma| on the expression, which joins two fully antisymmetrized gamma matrix products, and distributing factors over sums by  \verb|distribute|, we arrive at the below result:

\begin{verbatim}
join_gamma(_);
distribute(_);
\end{verbatim}

\begin{equation}
	\begin{array}{l}
\Gamma^{i j k l m} \Gamma_{m}+\Gamma^{i j l} \Gamma^{k}-\Gamma^{i j k} \Gamma^{l}+ \Gamma^{i l m} \Gamma_{m} \delta^{j k}-\Gamma^{i k m} \Gamma_{m} \delta^{j l}-\Gamma^{j l m} \Gamma_{m} \delta^{i k}+\Gamma^{j k m} \Gamma_{m} \delta^{i l}+\\ \Gamma^{i} \Gamma^{k} \delta^{j l}- \Gamma^{i} \Gamma^{l} \delta^{j k}-\Gamma^{m} \Gamma_{m} \delta^{i k} \delta^{j l}+\Gamma^{m} \Gamma_{m} \delta^{i l} \delta^{j k}+\Gamma^{j} \Gamma^{l} \delta^{i k}-\Gamma^{j} \Gamma^{k} \delta^{i l}
	\end{array}
\end{equation}

By repeating the above procedure once more:
\begin{verbatim}
join_gamma(_);
distribute(_);
\end{verbatim}

\begin{equation}
	\begin{array}{l}
2\Gamma^{i j k l}+4\Gamma^{i l} \delta^{j k}-4\Gamma^{j l} \delta^{i k}-4\Gamma^{i k} \delta^{j l}+4\Gamma^{j k} \delta^{i l}-6\delta^{i k} \delta^{j l}+6\delta^{i l} \delta^{j k}
	\end{array}
\end{equation}

Now, we perform a substitution for the term $\Gamma^{i j k l}$:
\begin{verbatim}
substitute(_, $\Gamma^{i j k l} -> \Gamma^{k l}\Gamma^{i j}
+\Gamma^{j k}\delta^{i l}-\Gamma^{i k}\delta^{j l}
-\Gamma^{j l}\delta^{i k}+\Gamma^{i l}\delta^{j k}
-\delta^{i l}\delta^{j k}+\delta^{i k}\delta^{j l}$ );
\end{verbatim}

Finally, one can arrive at the below useful identity:
\begin{equation}
	\begin{array}{l}
{}\Gamma^{i j m} \Gamma^{k l} \Gamma_{m}=
2\Gamma^{k l} \Gamma^{i j}+6\Gamma^{j k} \delta^{i l}-6\Gamma^{i k} \delta^{j l}-6\Gamma^{j l} \delta^{i k}+6\Gamma^{i l} \delta^{j k}+4\delta^{i l} \delta^{j k}-4\delta^{i k} \delta^{j l}
	\end{array}
\end{equation}

The above substitution is obtained using the relation from the below manipulation:
\begin{verbatim}
ex2:= \Gamma^{k l}\Gamma^{i j};
\end{verbatim}

\begin{equation}
	\begin{array}{l}
\Gamma^{k l} \Gamma^{i j}
	\end{array}
\end{equation}

\begin{verbatim}
join_gamma(_);
\end{verbatim}

\begin{equation}
	\begin{array}{l}
\Gamma^{i j k l}-\Gamma^{j k} \delta^{i l}+\Gamma^{i k} \delta^{j l}+\Gamma^{j l} \delta^{i k}-\Gamma^{i l} \delta^{j k}+\delta^{i l} \delta^{j k}-\delta^{i k} \delta^{j l}
	\end{array}
\end{equation}

On a product of four spinors, one can apply a Fierz transformation. Here, in eleven-dimensional Clifford algebra, we define our spinors as follows:

\begin{verbatim}
{m,n,p,q,r,s}::Indices;
{m,n,p,q,r,s}::Integer(0..10);
\Gamma{#}::GammaMatrix;
\bar{#}::DiracBar;
{\epsilon, \lambda, \chi, \psi}::Spinor;
\end{verbatim}

Consider the following expression of the spinor product:
\begin{verbatim}
ex:=\bar{\epsilon} \Gamma_{m} \psi \bar{\chi} \Gamma^{m} \lambda;
\end{verbatim}

\begin{equation}
	\begin{array}{l}
{}\bar{\epsilon} \Gamma_{m} \psi \bar{\chi} \Gamma^{m} \lambda
	\end{array}
\end{equation}
Now, we apply a Fierz transformation, with the desired order as follows:

\begin{verbatim}
fierz(_, $\epsilon, \lambda, \chi, \psi$);
\end{verbatim}

\begin{equation}
	\begin{array}{l}
{} - \frac{1}{32}\bar{\epsilon} \Gamma_{m} \Gamma^{m} \lambda \bar{\chi} \psi - \frac{1}{32}\bar{\epsilon} \Gamma_{m} \Gamma^{n} \Gamma^{m} \lambda \bar{\chi} \Gamma^{n} \psi - \frac{1}{64}\bar{\epsilon} \Gamma_{m} \Gamma^{n p} \Gamma^{m} \lambda \bar{\chi} \Gamma^{p n} \psi \\- \frac{1}{192}\bar{\epsilon} \Gamma_{m} \Gamma^{n p q} \Gamma^{m} \lambda \bar{\chi} \Gamma^{q p n} \psi - \frac{1}{768}\bar{\epsilon} \Gamma_{m} \Gamma^{n p q r} \Gamma^{m} \lambda \bar{\chi} \Gamma^{r q p n} \psi - \frac{1}{3840}\bar{\epsilon} \Gamma_{m} \Gamma^{n p q r s} \Gamma^{m} \lambda \bar{\chi} \Gamma^{s r q p n} \psi
	\end{array}
\end{equation}

\section{Superalgebra and Superfield Implementations}
Here, we first look at super-Poincar\'e algebra. We establish the vector and spinorial indices, and then, by declaring the antisymmetric and noncommuting features available in super-Poincar\'e generators, we simplify a commutation relation in this regard. Additionally, in the following subsection, by introducing the superspace formalism and supercharges in Cadabra, we shed light on the implementation of superfields, including scalar, chiral and vector superfields specifically for $N=1$ and $D=4$, to derive supersymmetric transformations of component fields.

First, we define the vector and spinor indices for the superalgebra generators as follows:
\begin{verbatim}
{a,b,c}::Indices(spinor);
{\mu,\nu,\rho,\sigma,\lambda,\kappa,\alpha,\beta,
\gamma,\xi}::Indices(vector);
{\mu,\nu,\rho,\sigma,\lambda,\kappa,\alpha,\beta,
\gamma,\xi}::Integer(0..3);
\delta{#}::KroneckerDelta;
\epsilon_{\mu\nu\lambda\rho}::EpsilonTensor(delta=\delta);
\end{verbatim}

At this stage, we declare the antisymmetric and noncommuting nature of our generators by attaching the properties \verb|AntiSymmetric|, \verb|SelfNonCommuting| and \verb|NonCommuting| to them.
\begin{verbatim}
{J_{\mu\nu},\sigma^{4}_{\mu\nu}} ::AntiSymmetric;
J_{\mu\nu}::SelfNonCommuting;
{ J_{\mu\nu}, P_{\mu}, W_{\mu} }::NonCommuting;
{ J_{\mu\nu}, W_{\mu},Q_{a}}::NonCommuting;
{J_{\mu\nu}, P_{\mu}, W_{\mu},Q_{a}}::Depends(\commutator{#});

\end{verbatim}

\begin{equation}
	\begin{array}{l}
\text{Property Indices(position=free) attached to~}{[}{a,~\discretionary{}{}{} b,~\discretionary{}{}{} c}{]}.\\

\text{Property Indices(position=free) attached to~}{[}{\mu,~\discretionary{}{}{} \nu,~\discretionary{}{}{} \rho,~\discretionary{}{}{} \sigma,~\discretionary{}{}{} \lambda,~\discretionary{}{}{} \kappa,~\discretionary{}{}{} \alpha,~\discretionary{}{}{} \beta,~\discretionary{}{}{} \gamma,~\discretionary{}{}{} \xi}{]}. \\

\text{Property Integer attached to~}{[}{\mu,~\discretionary{}{}{} \nu,~\discretionary{}{}{} \rho,~\discretionary{}{}{} \sigma,~\discretionary{}{}{} \lambda,~\discretionary{}{}{} \kappa,~\discretionary{}{}{} \alpha,~\discretionary{}{}{} \beta,~\discretionary{}{}{} \gamma,~\discretionary{}{}{} \xi}{]}.\\

\text{Property KroneckerDelta attached to~}\delta{(}{\#}{)}.\\

\text{Property EpsilonTensor attached to~}\epsilon_{\mu \nu \lambda \rho}.\\

\text{Property AntiSymmetric attached to~}\left[J_{\mu \nu},~ \sigma^{4}\,_{\mu \nu}\right].\\

\text{Property SelfNonCommuting attached to~}J_{\mu \nu}.\\

\text{Property NonCommuting attached to~}{[}{J_{\mu \nu},~\discretionary{}{}{} P_{\mu},~\discretionary{}{}{} W_{\mu}}{]}.\\

\text{Property NonCommuting attached to~}{[}{J_{\mu \nu},~\discretionary{}{}{} W_{\mu},~\discretionary{}{}{} Q_{a}}{]}.\\

\text{Property Depends attached to~}{[}{J_{\mu \nu},~\discretionary{}{}{} P_{\mu},~\discretionary{}{}{} W_{\mu},~\discretionary{}{}{} Q_{a}}{]}.
	\end{array}
\end{equation}

As mentioned in the previous section, simplification treatments are contained in the \verb|post_process| function, which is executed on every new input. Here, we update it to consider \verb|unwrap(ex)|, \verb|eliminate_kronecker(ex)|, \verb|canonicalise(ex)| and \verb|rename_dummies(ex)|:

\begin{verbatim}
def post_process(ex):
    unwrap(ex)
    eliminate_kronecker(ex)
    canonicalise(ex)
    rename_dummies(ex)
    collect_terms(ex)
\end{verbatim}

Now, we establish the commutator relations for $J_{\mu \nu}, Q_{a}$ and $P_{\mu}$ generators.
\begin{verbatim}
superpoincare:= { \commutator{J_{\mu\nu}}{Q_{a}} -> 
-\sigma^{4}_{\mu \nu} Q_{b},\commutator{P_{\mu}}{Q_{a}} -> 0 };
\end{verbatim}

\begin{equation}
	\begin{array}{l}
{}\left[{}\left[J_{\mu \nu}, Q_{a}\right]{} \rightarrow -\sigma^{4}\,_{\mu \nu} Q_{b},~ {}\left[P_{\mu}, Q_{a}\right]{} \rightarrow 0\right]
	\end{array}
\end{equation}

Below, we aim to compute the commutator between the square of the Pauli--Lubanski vector and supercharge $Q_{a}$:

\begin{verbatim}
Wsq:= \commutator{W_\mu W_\mu}{Q_{a}};
\end{verbatim}

\begin{equation}
	\begin{array}{l}
{}{}{[}{W_{\mu} W_{\mu}, \discretionary{}{}{}Q_{a}}{]}{}
	\end{array}
\end{equation}

We substitute the explicit definition of the Pauli--Lubanski vector into the commutator: 
\begin{verbatim}
substitute(_, $W_\mu -> 
1/2 \epsilon_{\mu\nu\lambda\rho} J_{\nu\lambda} P_\rho $);
\end{verbatim}

\begin{equation}
	\begin{array}{l}
{}\frac{1}{4}\epsilon_{\alpha \mu \nu \kappa} \epsilon_{\alpha \lambda \rho \sigma} {}{[}{J_{\mu \nu} P_{\lambda} J_{\rho \sigma} P_{\kappa}, \discretionary{}{}{}Q_{a}}{]}{}
	\end{array}
\end{equation}

One can replace the product of two epsilon tensors with a generalized delta by the below command:
\begin{verbatim}
epsilon_to_delta(_);
\end{verbatim}

\begin{equation}
	\begin{array}{l}
{}\frac{3}{2}\delta_{\mu \rho \nu \sigma \lambda \kappa} {}{[}{J_{\mu \nu} P_{\lambda} J_{\rho \sigma} P_{\kappa}, \discretionary{}{}{}Q_{a}}{]}{}
	\end{array}
\end{equation}

Now, we expand the generalized delta into standard two-index Kronecker deltas:
\begin{verbatim}
expand_delta(_);
\end{verbatim}

\begin{equation}
	\begin{array}{l}
\frac{3}{2}\left(\frac{1}{6}\delta_{\lambda \kappa} \delta_{\mu \rho} \delta_{\nu \sigma} - \frac{1}{6}\delta_{\lambda \nu} \delta_{\kappa \sigma} \delta_{\mu \rho} - \frac{1}{6}\delta_{\lambda \kappa} \delta_{\mu \sigma} \delta_{\nu \rho}+\frac{1}{6}\delta_{\lambda \mu} \delta_{\kappa \sigma} \delta_{\nu \rho}+\frac{1}{6}\delta_{\lambda \nu} \delta_{\kappa \rho} \delta_{\mu \sigma} 
- \frac{1}{6}\delta_{\lambda \mu} \delta_{\kappa \rho} \delta_{\nu \sigma}\right)\\ {}\left[J_{\mu \nu} P_{\lambda} J_{\rho \sigma} P_{\kappa}, Q_{a}\right]
	\end{array}
\end{equation}

By applying \verb|product_rule|, we can simplify the commutator in the above:
\begin{verbatim}
product_rule(_);
\end{verbatim}

\begin{equation}
	\begin{array}{l}
{}\frac{3}{2}\left(\frac{1}{6}\delta_{\lambda \kappa} \delta_{\mu \rho} \delta_{\nu \sigma} - \frac{1}{6}\delta_{\lambda \sigma} \delta_{\kappa \nu} \delta_{\mu \rho} - \frac{1}{6}\delta_{\lambda \kappa} \delta_{\mu \sigma} \delta_{\nu \rho}+\frac{1}{6}\delta_{\lambda \sigma} \delta_{\kappa \mu} \delta_{\nu \rho}+\frac{1}{6}\delta_{\lambda \rho} \delta_{\kappa \nu} \delta_{\mu \sigma} - \frac{1}{6}\delta_{\lambda \rho} \delta_{\kappa \mu} \delta_{\nu \sigma}\right)\\ \left({}\left[J_{\mu \nu}, Q_{a}\right]{} P_{\lambda} J_{\rho \sigma} P_{\kappa}+J_{\mu \nu} {}\left[P_{\kappa}, Q_{a}\right]{} J_{\rho \sigma} P_{\lambda}+J_{\mu \nu} P_{\lambda} {}\left[J_{\rho \sigma}, Q_{a}\right]{} P_{\kappa}+J_{\mu \nu} P_{\kappa} J_{\rho \sigma} {}\left[P_{\lambda}, Q_{a}\right]{}\right)

	\end{array}
\end{equation}

At this level, one can insert the super-Poincar\'e commutators defined previously into the expression and obtain: 
\begin{verbatim}
substitute(_, superpoincare);
\end{verbatim}

\begin{equation}
	\begin{array}{l}
{}\frac{3}{2}\left(\frac{1}{6}\delta_{\lambda \kappa} \delta_{\mu \rho} \delta_{\nu \sigma} - \frac{1}{6}\delta_{\lambda \nu} \delta_{\kappa \sigma} \delta_{\mu \rho} - \frac{1}{6}\delta_{\lambda \kappa} \delta_{\mu \sigma} \delta_{\nu \rho}+\frac{1}{6}\delta_{\lambda \nu} \delta_{\kappa \rho} \delta_{\mu \sigma}+\frac{1}{6}\delta_{\lambda \mu} \delta_{\kappa \sigma} \delta_{\nu \rho} - \frac{1}{6}\delta_{\lambda \mu} \delta_{\kappa \rho} \delta_{\nu \sigma}\right)\\ \left(-\sigma^{4}\,_{\rho \sigma} Q_{b} P_{\lambda} J_{\mu \nu} P_{\kappa}-J_{\rho \sigma} P_{\lambda} \sigma^{4}\,_{\mu \nu} Q_{b} P_{\kappa}\right)
	\end{array}
\end{equation}

Finally, we can distribute factors over sums and obtain a non-vanishing result:
\begin{verbatim}
distribute(_);
\end{verbatim}

\begin{equation}
	\begin{array}{l}
{} - \frac{1}{2}\sigma^{4}\,_{\mu \nu} Q_{b} P_{\rho} J_{\mu \nu} P_{\rho} - \frac{1}{2}J_{\mu \nu} P_{\rho} \sigma^{4}\,_{\mu \nu} Q_{b} P_{\rho}-\sigma^{4}\,_{\rho \mu} Q_{b} P_{\rho} J_{\mu \nu} P_{\nu}-J_{\mu \nu} P_{\mu} \sigma^{4}\,_{\nu \rho} Q_{b} P_{\rho}
	\end{array}
\end{equation}

\subsection{Scalar, Chiral and Vector Superfield Implementations}

In this section, we first define a general Lorentz scalar superfield in terms of its power series expansion.
Then, by defining the derivative and antiderivative operators (for Grassmann variables) and establishing the related anticommuting properties, we aim to derive supersymmetric transformations of component fields. We achieve this by first defining the infinitesimal supersymmetry variation and applying it to the superfield with \verb|substitute| and further simplifications via \verb|distribute|, \verb|product_rule| and \verb|unwrap|. At the final stage, we sort the coefficients based on the same powers of $\theta$ in the superfield expansion to read the SUSY transformation of the component field. We carry this out via hard-coded substitutions using \verb|take_match| and \verb|replace_match|. In a broader design, one can develop sorting rules to simplify general expressions containing Grassmann variables. A similar approach can be applied for the SUSY transformation of chiral and vector superfields.

Here, we start by defining a general scalar superfield \verb|Phi|:

\begin{verbatim}
\tbar{#}::LaTeXForm("\bar\theta").
\cbar{#}::LaTeXForm("\bar\chi").
\lambar{#}::LaTeXForm("\bar\lambda").
\aldot{#}::LaTeXForm("\dot\alpha").
\betdot{#}::LaTeXForm("\dot\beta").
\end{verbatim}

\begin{verbatim}
Superfield:= Phi-> (f(x)+\theta^{\beta} \phi_{\beta}+ \tbar_{\betdot} 
\cbar^{\betdot}+\indexbracket{\theta\theta}m(x)
+\indexbracket{\tbar\tbar}n(x)
+\theta^{\beta}\sigma_{\beta \betdot}^{\nu}\tbar^{\betdot} V_{\nu}
+\indexbracket{\theta\theta}\tbar_{\betdot}\lambar^{\betdot}
+\indexbracket{\tbar\tbar}\theta^{\beta}\psi_{\beta}
+\indexbracket{\theta\theta}\indexbracket{\tbar\tbar}d(x));
\end{verbatim}

\begin{equation}
	\begin{array}{l}
		Phi \rightarrow f{(}{x}{)}+\theta^{\beta} \phi_{\beta}+\bar\theta_{\dot\beta} \bar\chi^{\dot\beta}+{(}{\theta \theta}{)} m{(}{x}{)}+{(}{\bar\theta \bar\theta}{)} n{(}{x}{)}+\theta^{\beta} \sigma_{\beta \dot\beta}\,^{\nu} \bar\theta^{\dot\beta} V_{\nu}+{(}{\theta \theta}{)} \bar\theta_{\dot\beta} \bar\lambda^{\dot\beta}+{(}{\bar\theta \bar\theta}{)} \theta^{\beta} \psi_{\beta}\\+{(}{\theta \theta}{)} {(}{\bar\theta \bar\theta}{)} d{(}{x}{)}
	\end{array}
\end{equation}

\begin{verbatim}

\parbar{#}::LaTeXForm("\bar\partial").
\partialmu{#}::LaTeXForm("\hat\partial").
\epsbar{#}::LaTeXForm("\bar\epsilon").

{\partial{#},\parbar{#},\partialmu{#}}::PartialDerivative;


{\theta^{#},\tbar^{#},\epsilon^{#},\epsbar_{#}}::AntiCommuting;
{\theta^{\alpha},\tbar^{\aldot}}::SelfAntiCommuting;
\partial{#}::ImplicitIndex(\partial{#}_{\theta^{#}});
\parbar{#}::ImplicitIndex(\parbar{#}_{\tbar^{#}});

susy:=\epsilon^{\alpha} \partial_{\alpha}{Phi}+\epsbar_{\aldot}
\parbar^{\aldot}{Phi}
+\indexbracket{i\theta\sigma^{\mu}\epsbar} \partialmu_{\mu}{Phi}
-\indexbracket{i\epsilon\sigma^{\mu}\tbar} \partialmu_{\mu}{Phi};

\delta_{\alpha}^{\beta}::KroneckerDelta.
\delta^{\aldot}_{\betdot}::KroneckerDelta.
superderivative:={\partial_{\alpha}{\theta^{\beta}}->\delta_{\alpha}^{\beta},
	\parbar^{\aldot}{\tbar_{\betdot}}->\delta^{\aldot}_{\betdot},
	\parbar^{\aldot}{\tbar^{\betdot}}->-e^{\aldot \betdot},
	\partial_{\alpha}{\indexbracket{\theta\theta}}->2\theta_{\alpha},
	\parbar^{\aldot}{\indexbracket{\tbar\tbar}}->2\tbar^{\aldot}};
\end{verbatim}

\begin{equation}
	\begin{array}{l}
	
	\text{Property PartialDerivative attached to~}{[}\partial{\#},\bar\partial{\#},\hat\partial{\#}{]}.\\
	
	{}\text{Property AntiCommuting attached to~}{[}{\theta^{\#},~\discretionary{}{}{} \bar\theta^{\#}}{,\epsilon^{\#},\bar\epsilon_{\#}}].\\
{}\text{Property SelfAntiCommuting attached to~}{[}{\theta^{\alpha},~\discretionary{}{}{} \bar\theta^{\dot\alpha}}{]}.\\
{}\text{Property ImplicitIndex attached to~}\partial{(}{\#}{)}.\\
{}\text{Property ImplicitIndex attached to~}\bar\partial{(}{\#}{)}.\\
	
	\epsilon^{\alpha} \partial_{\alpha}{(}{Phi}{)}+\bar\epsilon_{\dot\alpha} \bar\partial^{\dot\alpha}{(}{Phi}{)}+{(}{i \theta \sigma^{\mu} \bar\epsilon}{)} \hat\partial_{\mu}{(}{Phi}{)}-{(}{i \epsilon \sigma^{\mu} \bar\theta}{)} \hat\partial_{\mu}{(}{Phi}{)}\\

{}{[}{\partial_{\alpha}{(}{\theta^{\beta}}{)} \rightarrow \delta_{\alpha}\,^{\beta},~\discretionary{}{}{} \bar\partial^{\dot\alpha}{(}{\bar\theta_{\dot\beta}}{)} \rightarrow \delta^{\dot\alpha}\,_{\dot\beta},~\discretionary{}{}{} \bar\partial^{\dot\alpha}{(}{\bar\theta^{\dot\beta}}{)} \rightarrow -e^{\dot\alpha \dot\beta},~\discretionary{}{}{} \partial_{\alpha}{(}{{(}{\theta \theta}{)}}{)} \rightarrow 2\theta_{\alpha},~\discretionary{}{}{} \bar\partial^{\dot\alpha}{(}{{(}{\bar\theta \bar\theta}{)}}{)} \rightarrow 2\bar\theta^{\dot\alpha}}{]}
	\end{array}
\end{equation}

\begin{verbatim}
substitute(susy,Superfield);
\end{verbatim}

\begin{verbatim}
distribute(_)
product_rule(_)

{\theta^{#},\indexbracket{\theta\theta}}
::Depends(\partial{#});
{\tbar_{#},\indexbracket{\tbar\tbar}}
::Depends(\parbar{#});
{f(x),m(x),n(x),\phi_{\alpha},\cbar^{\aldot},V_{\nu},\lambar^{\betdot},
	\psi_{\beta}}::Depends(\partialmu{#});
\end{verbatim}

\begin{equation}
	\begin{array}{l}
\text{Property Depends attached to~}{[}{\theta^{\#},{(}{\theta \theta}{)}}{]}.\\

\text{Property Depends attached to~}{[}{\bar\theta_{\#} ,{(}{\bar\theta \bar\theta}{)}}{]}.\\

\text{Property Depends attached to~}{[}{f{(}{x}{)},~\discretionary{}{}{} m{(}{x}{)},~\discretionary{}{}{} n{(}{x}{)},~\discretionary{}{}{} \phi_{\alpha},~\discretionary{}{}{} \bar\chi^{\dot\alpha},~\discretionary{}{}{} V_{\nu},~\discretionary{}{}{} \bar\lambda^{\dot\beta},~\discretionary{}{}{} \psi_{\beta}}{]}.
	\end{array}
\end{equation}

\begin{verbatim}
unwrap(_);
\end{verbatim}

\begin{equation}
	\begin{array}{l}
\epsilon^{\alpha} \partial_{\alpha}{\theta^{\beta}} \phi_{\beta}+\epsilon^{\alpha} \partial_{\alpha}{(}{{(}{\theta \theta}{)}}{)} m{(}{x}{)}+\epsilon^{\alpha} \partial_{\alpha}{\theta^{\beta}} \sigma_{\beta \dot\beta}\,^{\nu} \bar\theta^{\dot\beta} V_{\nu}+\epsilon^{\alpha} \partial_{\alpha}{(}{{(}{\theta \theta}{)}}{)} \bar\theta_{\dot\beta} \bar\lambda^{\dot\beta}+\epsilon^{\alpha} {(}{\bar\theta \bar\theta}{)} \partial_{\alpha}{\theta^{\beta}} \psi_{\beta}+\\\epsilon^{\alpha} \partial_{\alpha}{(}{{(}{\theta \theta}{)}}{)} {(}{\bar\theta \bar\theta}{)} d{(}{x}{)}+\bar\epsilon_{\dot\alpha} \bar\partial^{\dot\alpha}{\bar\theta_{\dot\beta}} \bar\chi^{\dot\beta}+\bar\epsilon_{\dot\alpha} \bar\partial^{\dot\alpha}{(}{{(}{\bar\theta \bar\theta}{)}}{)} n{(}{x}{)}-\bar\epsilon_{\dot\alpha} \theta^{\beta} \sigma_{\beta \dot\beta}\,^{\nu} \bar\partial^{\dot\alpha}{\bar\theta^{\dot\beta}} V_{\nu}+\bar\epsilon_{\dot\alpha} {(}{\theta \theta}{)} \bar\partial^{\dot\alpha}{\bar\theta_{\dot\beta}} \bar\lambda^{\dot\beta}+\\\bar\epsilon_{\dot\alpha} \bar\partial^{\dot\alpha}{(}{{(}{\bar\theta \bar\theta}{)}}{)} \theta^{\beta} \psi_{\beta}+\bar\epsilon_{\dot\alpha} {(}{\theta \theta}{)} \bar\partial^{\dot\alpha}{(}{{(}{\bar\theta \bar\theta}{)}}{)} d{(}{x}{)}+{(}{i \theta \sigma^{\mu} \bar\epsilon}{)} \hat\partial_{\mu}{f{(}{x}{)}}+{(}{i \theta \sigma^{\mu} \bar\epsilon}{)} \theta^{\beta} \hat\partial_{\mu}{\phi_{\beta}}+{(}{i \theta \sigma^{\mu} \bar\epsilon}{)} \bar\theta_{\dot\beta} \hat\partial_{\mu}{\bar\chi^{\dot\beta}}+\\{(}{i \theta \sigma^{\mu} \bar\epsilon}{)} {(}{\theta \theta}{)} \hat\partial_{\mu}{m{(}{x}{)}}+{(}{i \theta \sigma^{\mu} \bar\epsilon}{)} {(}{\bar\theta \bar\theta}{)} \hat\partial_{\mu}{n{(}{x}{)}}+{(}{i \theta \sigma^{\mu} \bar\epsilon}{)} \theta^{\beta} \sigma_{\beta \dot\beta}\,^{\nu} \bar\theta^{\dot\beta} \hat\partial_{\mu}{V_{\nu}}+{(}{i \theta \sigma^{\mu} \bar\epsilon}{)} {(}{\theta \theta}{)} \bar\theta_{\dot\beta} \hat\partial_{\mu}{\bar\lambda^{\dot\beta}}%
\\+{(}{i \theta \sigma^{\mu} \bar\epsilon}{)} {(}{\bar\theta \bar\theta}{)} \theta^{\beta} \hat\partial_{\mu}{\psi_{\beta}}-{(}{i \epsilon \sigma^{\mu} \bar\theta}{)} \hat\partial_{\mu}{f{(}{x}{)}}-{(}{i \epsilon \sigma^{\mu} \bar\theta}{)} \theta^{\beta} \hat\partial_{\mu}{\phi_{\beta}}-{(}{i \epsilon \sigma^{\mu} \bar\theta}{)} \bar\theta_{\dot\beta} \hat\partial_{\mu}{\bar\chi^{\dot\beta}}-{(}{i \epsilon \sigma^{\mu} \bar\theta}{)} {(}{\theta \theta}{)} \hat\partial_{\mu}{m{(}{x}{)}}\\-{(}{i \epsilon \sigma^{\mu} \bar\theta}{)} {(}{\bar\theta \bar\theta}{)} \hat\partial_{\mu}{n{(}{x}{)}}-{(}{i \epsilon \sigma^{\mu} \bar\theta}{)} \theta^{\beta} \sigma_{\beta \dot\beta}\,^{\nu} \bar\theta^{\dot\beta} \hat\partial_{\mu}{V_{\nu}}-{(}{i \epsilon \sigma^{\mu} \bar\theta}{)} {(}{\theta \theta}{)} \bar\theta_{\dot\beta} \hat\partial_{\mu}{\bar\lambda^{\dot\beta}}-{(}{i \epsilon \sigma^{\mu} \bar\theta}{)} {(}{\bar\theta \bar\theta}{)} \theta^{\beta} \hat\partial_{\mu}{\psi_{\beta}}
	\end{array}
\end{equation}

\begin{verbatim}
take_match(_, $\indexbracket{i\theta\sigma^{\mu}\epsbar}
\indexbracket{\theta\theta} Q??$)
substitute(_, $A??+ B?? -> 0$)
replace_match(_)

take_match(_, $\indexbracket{i\epsilon\sigma^{\mu}\tbar}
\indexbracket{\tbar\tbar} Q??$)
substitute(_, $A??+ B?? -> 0$)
replace_match(_)

substitute(_,superderivative)
eliminate_kronecker(_);
\end{verbatim}

\begin{equation}
	\begin{array}{l}

\epsilon^{\beta} \phi_{\beta}+2\epsilon^{\alpha} \theta_{\alpha} m{(}{x}{)}+\epsilon^{\beta} \sigma_{\beta \dot\beta}\,^{\nu} \bar\theta^{\dot\beta} V_{\nu}+2\epsilon^{\alpha} \theta_{\alpha} \bar\theta_{\dot\beta} \bar\lambda^{\dot\beta}+\epsilon^{\beta} {(}{\bar\theta \bar\theta}{)} \psi_{\beta}+2\epsilon^{\alpha} \theta_{\alpha} {(}{\bar\theta \bar\theta}{)} d{(}{x}{)}+\bar\epsilon_{\dot\beta} \bar\chi^{\dot\beta}\\+2\bar\epsilon_{\dot\alpha} \bar\theta^{\dot\alpha} n{(}{x}{)}+\bar\epsilon_{\dot\alpha} \theta^{\beta} \sigma_{\beta \dot\beta}\,^{\nu} e^{\dot\alpha \dot\beta}V_{\nu}+\bar\epsilon_{\dot\beta} {(}{\theta \theta}{)} \bar\lambda^{\dot\beta}+2\bar\epsilon_{\dot\alpha} \bar\theta^{\dot\alpha} \theta^{\beta} \psi_{\beta}+2\bar\epsilon_{\dot\alpha} {(}{\theta \theta}{)} \bar\theta^{\dot\alpha} d{(}{x}{)}\\+{(}{i \theta \sigma^{\mu} \bar\epsilon}{)} \hat\partial_{\mu}{f{(}{x}{)}}+{(}{i \theta \sigma^{\mu} \bar\epsilon}{)} \theta^{\beta} \hat\partial_{\mu}{\phi_{\beta}}+{(}{i \theta \sigma^{\mu} \bar\epsilon}{)} \bar\theta_{\dot\beta} \hat\partial_{\mu}{\bar\chi^{\dot\beta}}+{(}{i \theta \sigma^{\mu} \bar\epsilon}{)} {(}{\bar\theta \bar\theta}{)} \hat\partial_{\mu}{n{(}{x}{)}}\\+{(}{i \theta \sigma^{\mu} \bar\epsilon}{)} \theta^{\beta} \sigma_{\beta \dot\beta}\,^{\nu} \bar\theta^{\dot\beta} \hat\partial_{\mu}{V_{\nu}}+{(}{i \theta \sigma^{\mu} \bar\epsilon}{)} {(}{\bar\theta \bar\theta}{)} \theta^{\beta} \hat\partial_{\mu}{\psi_{\beta}}-{(}{i \epsilon \sigma^{\mu} \bar\theta}{)} \hat\partial_{\mu}{f{(}{x}{)}}%
\\-{(}{i \epsilon \sigma^{\mu} \bar\theta}{)} \theta^{\beta} \hat\partial_{\mu}{\phi_{\beta}}-{(}{i \epsilon \sigma^{\mu} \bar\theta}{)} \bar\theta_{\dot\beta} \hat\partial_{\mu}{\bar\chi^{\dot\beta}}-{(}{i \epsilon \sigma^{\mu} \bar\theta}{)} {(}{\theta \theta}{)} \hat\partial_{\mu}{m{(}{x}{)}}\\-{(}{i \epsilon \sigma^{\mu} \bar\theta}{)} \theta^{\beta} \sigma_{\beta \dot\beta}\,^{\nu} \bar\theta^{\dot\beta} \hat\partial_{\mu}{V_{\nu}}-{(}{i \epsilon \sigma^{\mu} \bar\theta}{)} {(}{\theta \theta}{)} \bar\theta_{\dot\beta} \hat\partial_{\mu}{\bar\lambda^{\dot\beta}}
	\end{array}
\end{equation}

\begin{verbatim}

take_match(_, $\theta^{\beta}\partialmu_{\mu}{\phi_{\beta}} Q??$)
take_match(_, $\indexbracket{i\theta\sigma^{\mu}\epsbar} Q??$)
substitute(_,$A?? B?? C??->\indexbracket{\theta\theta}
\indexbracket{-i/2 \partialmu_{\nu}{\phi}\sigma^{\nu}\epsbar}$)
replace_match(_)
replace_match(_)

take_match(_, $\tbar_{\betdot}\partialmu_{\mu}{\cbar^{\betdot}} Q??$)
take_match(_, $\indexbracket{i\epsilon\sigma^{\mu}\tbar} Q??$)
substitute(_,$A?? B?? C??-> -\indexbracket{\tbar\tbar}
\indexbracket{i/2 \epsilon\sigma^{\nu}\partialmu_{\nu}{\cbar}}$)
replace_match(_)
replace_match(_)

take_match(_, $\indexbracket{i\theta\sigma^{\mu}\epsbar} 
\partialmu_{\mu}{V_{\nu}} Q??$)
substitute(_,$A?? B?? C?? D?? E??->
i/2 \indexbracket{\theta\theta}\indexbracket{\tbar\epsbar}
\partialmu^{\mu}{V_{\mu}}$)
replace_match(_)

take_match(_, $\partialmu_{\mu}{m(x)} Q??$)
substitute(_,$\indexbracket{i\epsilon\sigma^{\mu}\tbar}
\indexbracket{\theta\theta}->-i \indexbracket{\theta\theta}\tbar_{\betdot}
\indexbracket{\epsilon\sigma^{\mu}e}^{\betdot}}$)
replace_match(_)

take_match(_, $\partialmu_{\mu}{V_{\nu}} Q??$)
substitute(_,$K?? L?? M?? N?? O??->
i/2 \indexbracket{\tbar\tbar}\indexbracket{\theta\epsilon}
\partialmu^{\mu}{V_{\mu}}$)
replace_match(_)

take_match(_, $\partialmu_{\mu}{\psi_{\beta}} Q??$)
substitute(_,$K?? L?? M?? N??  ->i/2 \indexbracket{\theta\theta}
\indexbracket{\tbar\tbar}\partialmu_{\nu}{\psi}\sigma^{\nu}\epsbar$)
replace_match(_)

take_match(_, $\partialmu_{\mu}{\lambar^{\betdot}} Q??$)
substitute(_,$K?? L?? M?? N??  ->
-i/2 \indexbracket{\theta\theta}\indexbracket{\tbar\tbar}
\epsilon\sigma^{\nu}\partialmu_{\nu}{\lambar}$)
replace_match(_)

take_match(_, $\tbar_{\betdot}\lambar^{\betdot} Q??$)
substitute(_,$K?? L?? M?? N?? ->
1/2 \indexbracket{\theta\sigma^{\mu}\tbar}
\indexbracket{\epsilon\sigma_{\mu}\lambar}$)
replace_match(_)

take_match(_, $\theta^{\beta}\psi_{\beta} Q??$)
substitute(_,$K?? L?? M?? N?? ->
1/2 \indexbracket{\theta\sigma^{\mu}\tbar}
\indexbracket{\psi\sigma_{\mu}\epsbar}$)
replace_match(_)

take_match(_, $\partialmu_{\mu}{\phi_{\beta}} Q??$)
substitute(_,$K?? L?? M?? ->
-i/2 \indexbracket{\theta\sigma^{\nu}\tbar}
\indexbracket{\epsilon\partialmu_{\nu}{\phi}$)
replace_match(_)

take_match(_, $\tbar_{\betdot}\partialmu_{\mu}{\cbar^{\beta}} Q??$)
substitute(_,$K?? L?? M?? ->
-i/2 \indexbracket{\theta\sigma_{\nu}\tbar}
\indexbracket{\partialmu^{\nu}{\cbar}\epsbar}$)
replace_match(_);
\end{verbatim}

\begin{equation}
	\begin{array}{l}
\epsilon^{\beta} \phi_{\beta}+2\epsilon^{\alpha} \theta_{\alpha} m{(}{x}{)}+\epsilon^{\beta} \sigma_{\beta \dot\beta}\,^{\nu} \bar\theta^{\dot\beta} V_{\nu}+\epsilon^{\beta} {(}{\bar\theta \bar\theta}{)} \psi_{\beta}+2\epsilon^{\alpha} \theta_{\alpha} {(}{\bar\theta \bar\theta}{)} d{(}{x}{)}+\bar\epsilon_{\dot\beta} \bar\chi^{\dot\beta}+2\bar\epsilon_{\dot\alpha} \bar\theta^{\dot\alpha} n{(}{x}{)}\\+\bar\epsilon_{\dot\alpha} \theta^{\beta} \sigma_{\beta \dot\beta}\,^{\nu} e^{\dot\alpha \dot\beta} V_{\nu}+\bar\epsilon_{\dot\beta} {(}{\theta \theta}{)} \bar\lambda^{\dot\beta}+2\bar\epsilon_{\dot\alpha} {(}{\theta \theta}{)} \bar\theta^{\dot\alpha} d{(}{x}{)}+{(}{i \theta \sigma^{\mu} \bar\epsilon}{)} \hat\partial_{\mu}{f{(}{x}{)}}+{(}{i \theta \sigma^{\mu} \bar\epsilon}{)} {(}{\bar\theta \bar\theta}{)} \hat\partial_{\mu}{n{(}{x}{)}}\\-{(}{i \epsilon \sigma^{\mu} \bar\theta}{)} \hat\partial_{\mu}{f{(}{x}{)}}+{(}{\theta \theta}{)} {(}{ - \frac{1}{2}i \hat\partial_{\nu}{\phi} \sigma^{\nu} \bar\epsilon}{)}+{(}{\bar\theta \bar\theta}{)} {(}{\frac{1}{2}i \epsilon \sigma^{\nu} \hat\partial_{\nu}{\bar\chi}}{)}+\frac{1}{2}i {(}{\theta \theta}{)} {(}{\bar\theta \bar\epsilon}{)} \hat\partial^{\mu}{V_{\mu}}\\+i {(}{\theta \theta}{)} \bar\theta_{\dot\beta} {(}{\epsilon \sigma^{\mu} e}{)}\,^{\dot\beta} \hat\partial_{\mu}{m{(}{x}{)}} - \frac{1}{2}i {(}{\bar\theta \bar\theta}{)} {(}{\theta \epsilon}{)} \hat\partial^{\mu}{V_{\mu}}+\frac{1}{2}i {(}{\theta \theta}{)} {(}{\bar\theta \bar\theta}{)} \hat\partial_{\nu}{\psi} \sigma^{\nu} \bar\epsilon%
\\+\frac{1}{2}i {(}{\theta \theta}{)} {(}{\bar\theta \bar\theta}{)} \epsilon \sigma^{\nu} \hat\partial_{\nu}{\bar\lambda}+{(}{\theta \sigma^{\mu} \bar\theta}{)} {(}{\epsilon \sigma_{\mu} \bar\lambda}{)}+{(}{\theta \sigma^{\mu} \bar\theta}{)} {(}{\psi \sigma_{\mu} \bar\epsilon}{)}\\+\frac{1}{2}i {(}{\theta \sigma^{\nu} \bar\theta}{)} {(}{\epsilon \hat\partial_{\nu}{\phi}}{)} - \frac{1}{2}i {(}{\theta \sigma_{\nu} \bar\theta}{)} {(}{\hat\partial^{\nu}{\bar\chi} \bar\epsilon}{)}
	\end{array}
\end{equation}

Now, one can easily read off the supersymmetry transformation of the component fields from the above expression.

One may find that the various hard-coded substitution routines at specific terms that we used to achieve the canonical form for $\theta$ expansions make the computation procedure more difficult.  As mentioned previously, a broader algorithm would be more feasible to sort these Grassmann expressions at least to a better stopping point. For example, in the explicit indices, we can have:

\begin{verbatim}

{\tbar^{#}, \theta^{#}}::SelfAntiCommuting;
{\tbar^{#},\epsbar^{#}, \theta^{#},\epsilon^{#},\psi_{#}}::AntiCommuting;
{\tbar^{#},\theta^{#},\epsilon^{#}}::SortOrder;
\partial{#}::PartialDerivative;
ex:=\epsilon^{\alpha} M^{\mu}_{\alpha \aldot} \tbar^{\aldot} 
\theta^{\beta} M^{\nu}_{\beta \betdot} \tbar^{\betdot} 
\partial_{\mu}{V_{\nu}} + 
\theta^{\beta} M^{\mu}_{\beta \aldot} \epsbar^{\aldot} 
\tbar_{\betdot}\tbar^{\betdot} \theta^{\alpha} \partial_{\mu}{\psi_{\alpha}};
\end{verbatim}

\begin{equation}
	\begin{array}{l}
{}\text{Property SelfAntiCommuting attached to~}{[}{\bar\theta^{\#},~\discretionary{}{}{} \theta^{\#}}{]}.\\
{}\text{Property AntiCommuting attached to~}{[}{\bar\theta^{\#},~\discretionary{}{}{} \bar\epsilon^{\#},~\discretionary{}{}{} \theta^{\#},~\discretionary{}{}{} \epsilon^{\#},~\discretionary{}{}{} \psi_{\#}}{]}.\\
{}\text{Property SortOrder attached to~}{[}{\bar\theta^{\#},~\discretionary{}{}{} \theta^{\#},~\discretionary{}{}{} \epsilon^{\#}}{]}.\\
{}\text{Property PartialDerivative attached to~}\partial{\#}.\\
{}\epsilon^{\alpha} M^{\mu}\,_{\alpha \dot\alpha} \bar\theta^{\dot\alpha} \theta^{\beta} M^{\nu}\,_{\beta \dot\beta} \bar\theta^{\dot\beta} \partial_{\mu}{V_{\nu}}+\theta^{\beta} M^{\mu}\,_{\beta \dot\alpha} \bar\epsilon^{\dot\alpha} \bar\theta_{\dot\beta} \bar\theta^{\dot\beta} \theta^{\alpha} \partial_{\mu}{\psi_{\alpha}}

	\end{array}
\end{equation}
Here, $M^{\mu}$ is the Pauli matrices and the anticommutativity of our spinors is indicated using the \verb|AntiCommuting| property. We also define our preferred order using \verb|SortOrder|. Now, we apply \verb|sort_product| to sort factors in our Grassmann expression defined above:

\begin{verbatim}
sort_product(_);
\end{verbatim}

\begin{equation}
	\begin{array}{l}
{}M^{\mu}\,_{\alpha \dot\alpha} M^{\nu}\,_{\beta \dot\beta} \bar\theta^{\dot\alpha} \bar\theta^{\dot\beta} \theta^{\beta} \epsilon^{\alpha} \partial_{\mu}{V_{\nu}}+M^{\mu}\,_{\beta \dot\alpha} \bar\epsilon^{\dot\alpha} \partial_{\mu}{\psi_{\alpha}} \bar\theta_{\dot\beta} \bar\theta^{\dot\beta} \theta^{\alpha} \theta^{\beta}
	\end{array}
\end{equation}

After this stage, one can use the relation among the Grassmann variables for further simplifications. Finally, working with $\theta$-expansions can be cumbersome. One can handle things in a covariant way, using projection techniques to find component field expansions of supersymmetric actions. The implementation of projection operators via supercovariant derivatives in the current stage of Cadabra is a little tricky, but it is still possible to perform computations on them.

At the end, chiral and vector superfields can be implemented in Cadabra in the same way as general scalar superfields. Here, we quickly review chiral and vector superfields. We use these multiplets in the next section for the automation of supersymmetric Lagrangians.

A superfield $\Phi $ satisfying the constraint $\bar D \Phi=0$ is called a left chiral superfield, where $\bar D$ is a covariant derivative for the superfield. In terms of new variable $y^{\mu}=x^{\mu}+i\theta\sigma^{\mu}\bar\theta$, the superfield has the following power series expansion in $\theta$:

\begin{equation}
	\begin{array}{l}
		\Phi ({y^\mu },{\theta ^\alpha }) = A (y) + \sqrt 2 \theta \psi (y) + \theta \theta F(y)
	\end{array}
\end{equation}
where A(y) and F(y) are complex scalar fields, and $\psi(y)$ is a left-handed Weyl spinor.
By expanding the component fields in terms of variables $x,\theta,\bar\theta$, we have
\begin{equation}
	\begin{array}{l}
		\Phi ({x^\mu },{\theta ^\alpha },{{\bar \theta }^{\dot \alpha }}) = \phi (x) + \sqrt 2 \theta \psi (x) + \theta \theta F(x) + i\theta {\sigma ^\mu }\bar \theta {\partial _\mu }\phi (x)\\
		- \frac{i}{{\sqrt 2 }}(\theta \theta ){\partial _\mu }\psi (x){\sigma ^\mu }\bar \theta  - \frac{1}{4}(\theta \theta )(\bar \theta \bar \theta ){\partial _\mu }{\partial ^\mu }\phi (x)
	\end{array}
\end{equation}

Under supersymmetry transformation, we have
\begin{equation}
\delta \Phi  = i(\varepsilon Q + \bar \varepsilon \bar Q)\Phi 
\end{equation}

As we defined the differential operator representations of these supercharges, one can easily compute the SUSY transformation for the chiral field using the same approach as we used for a general scalar superfield.

A vector superfield $V({x^\mu },{\theta ^\alpha },{{\bar \theta }^{\dot \alpha }})$ is defined by the reality condition $V=V^{\dagger}$. In the Wess--Zumino gauge, a vector superfield is expressed as:

\begin{equation}
V_{W.Z}=(\theta\sigma^{\mu}\bar\theta)V_{\mu}+(\theta\theta)\bar\theta \bar\lambda(x)+(\bar\theta\bar\theta)\theta\lambda(x)+(\theta\theta)(\bar\theta\bar\theta)D(x)
\end{equation}
where $V_{\mu}(x)$ is a real vector field, $\lambda(x)$ is a complex spinor field and D(x) is a real scalar field.

In section 3, we choose a vector superfield to define the invariant kinetic part of the superspace Lagrangian density and to show how one can achieve simple automation for SUSY Lagrangian generation.

\section{Supersymmetric Lagrangian Implementations}
In an automated way, one can construct supersymmetric Lagrangians and actions and verify the various properties and structures of the related SUSY model directly within Cadabra. 

Here, we shed light on the potential capabilities of Cadabra in the automatic generation of supersymmetric Lagrangians, which can speed up verification tasks and facilitate the study of SUSY theories that are analytically computation-intensive. 
We do not address all of the details of Cadabra implementations and only elaborate on areas in which one can develop Cadabara's algorithms and functions to handle the automation.

Part of the supersymmetric Lagrangian describing the dynamics of different component fields in chiral and vector supermultiplets can be automated. The most general supersymmetry Lagrangian describing the dynamics between various multiplets can be constructed based on invariant kinetic and superpotential parts and gauge interactions:
 
\begin{equation}
	{\mathcal{L} = {\left. {{\Phi ^\dag }_i{e^{{q_j}{V^j}}}{\Phi _i}} \right|_{\theta \theta \bar \theta \bar \theta }}+{\left. {W[\Phi ]} \right|_{\theta \theta }} + {\left. {\bar W[{\Phi ^\dag }]} \right|_{\bar \theta \bar \theta }}+\frac{1}{{16}}{\left. {W_A^\alpha W_\alpha ^A} \right|_{\theta \theta }} + \frac{1}{{16}}{\left. {\bar W_{\dot \alpha }^A\bar W_A^{\dot \alpha }} \right|_{\bar \theta \bar \theta }}}
\end{equation}
 where $W[\Phi ]$ is a polynomial of the superfield $\Phi$ and is called the superpotential. $W_A^\alpha$ and $\bar W_{\dot \alpha }^A$ are our spinorial superfields. Based on the above structure, Lagrangian density generation can be automated by defining three main functions in Cadabra as the extractor of the ${\theta \theta \bar \theta \bar \theta }$, ${\theta \theta }$ and ${\bar \theta \bar \theta }$ components.

Here, for educational purposes, we show how to achieve simple automation for the gauge covariant kinetic term ( ${\theta \theta \bar \theta \bar \theta }$ component extractor). Similarly, one can develop ${\theta \theta}$ and ${\bar \theta \bar \theta }$ component extractors out of scalar superfields of $W_A^\alpha W_\alpha ^A$ and $\bar W_{\dot \alpha }^A\bar W_A^{\dot \alpha }$ and from superpotentials $W[\Phi ]$ and $\bar W[{\Phi ^\dag }]$.

We define a simple toy function called \verb|T2TBar2| to derive the invariant kinetic part of the SUSY  Lagrangian from a list of chiral superfields in the Wess--Zumino gauge. Here, for simplicity of the algorithm, we do not search for the component of the vector superfield and use preconfigured components of $V_{\mu}(x)$ , $\lambda (x)$ and $D(x)$.

\begin{verbatim}
def T2TBar2(SF):
    F_terms:=0;
    Scalar_terms:=0;
    Spinor_lambda:=0;
    Spinors_psi:=0;
    Total_Terms:=0;

    for element in SF["F"]:
        F_term:= \mid \indexbracket{@(element)} \mid**2.
        F_terms += F_term

    for element in SF["A"]:
        indexa = element.indices().__next__()
        Scalar_term:= q_{@(indexa)}D(x) \mid 
        \indexbracket{@(element)}\mid**2}  
        + (\partial_{\mu}{@(element)}\partial^{\mu}{@(element)^{\star}}
        +1/4 q_{@(indexa)}**2 V_{\mu}(x) V^{\mu}(x) \mid
        \indexbracket{@(element)}\mid**2.
        Scalar_terms += Scalar_term

        for psi in SF["\psi"]:
            indexpsi = psi.indices().__next__()
            if str(indexpsi)==str(indexa):
                Spinor_term1:= -1/\sqrt{2}} q_{@(indexa)} 
                \indexbracket{\lambda(x) @(psi)} @(element)^{\star}.
                Spinor_term2:= -1/\sqrt{2} q_{@(indexa)}\bar{\lambda(x)} 
                \bar{@(psi)} @(element).
            else:
                Spinor_term1:=0 ;
                Spinor_term2:=0 ;

            Spinor_lambda +=Spinor_term1+Spinor_term2

    for psi in SF["\psi"]:
        indexpsi = psi.indices().__next__()
        Spinor:= i \partial_{\mu}{\bar{@(psi)}}\bar{\sigma}^{\mu} 
        @(psi)-1/2 q_{@(indexpsi)} V(x)_{\mu}\bar{@(psi)}
        \bar{\sigma}^{\mu} @(psi).
        Spinors_psi += Spinor 
    Total_Terms=  F_terms + Scalar_terms + Spinor_lambda + Spinors_psi
    return Total_Terms

\end{verbatim}

The input is a simple list of field contents. For example, consider the case of two scalar superfields (chiral):

\begin{verbatim}
	SF:=[A_{1}(x),\psi_{1}(x),F_{1}(x),q_{1},A_{2}(x),\psi_{2}(x),
	F_{2}(x),q_{2},V_{\mu}(x),\lambda(x),D(x)];
\end{verbatim}

\begin{equation}	
	{[}{A_{1}{(}{x}{)},~\discretionary{}{}{} \psi_{1}{(}{x}{)},~\discretionary{}{}{} F_{1}{(}{x}{)},~\discretionary{}{}{} q_{1},~\discretionary{}{}{} A_{2}{(}{x}{)},~\discretionary{}{}{} \psi_{2}{(}{x}{)},~\discretionary{}{}{} F_{2}{(}{x}{)},~\discretionary{}{}{} q_{2},~\discretionary{}{}{} V_{\mu}{(}{x}{)},~\discretionary{}{}{} \lambda{(}{x}{)},~\discretionary{}{}{} D{(}{x}{)}}{]}
\end{equation}

\begin{verbatim}
T2TBar2(SF);
\end{verbatim}

\begin{equation}	
	\begin{array}{l}
{}\mid {(}{F_{1}{(}{x}{)}}{)} {\mid}^{2}+\mid {(}{F_{2}{(}{x}{)}}{)} {\mid}^{2}+  q_{1} D{(}{x}{)} \mid {(}{A_{1}{(}{x}{)}}{)} {\mid}^{2}+\partial_{\mu}{[}{A_{1}{(}{x}{)}}{]} \partial^{\mu}{[}{A_{1}{(}{x}{)}\,^{\star}}{]}+\\\frac{1}{4}{q_{1}}^{2} V_{\mu}{(}{x}{)} V^{\mu}{(}{x}{)} \mid {(}{A_{1}{(}{x}{)}}{)} {\mid}^{2}+q_{2} D{(}{x}{)} \mid {(}{A_{2}{(}{x}{)}}{)} {\mid}^{2}+\partial_{\mu}{[}{A_{2}{(}{x}{)}}{]} \partial^{\mu}{[}{A_{2}{(}{x}{)}\,^{\star}}{]}+\\\frac{1}{4}{q_{2}}^{2} V_{\mu}{(}{x}{)} V^{\mu}{(}{x}{)} \mid {(}{A_{2}{(}{x}{)}}{)} {\mid}^{2}-{2}^{ - \frac{1}{2}} q_{1} {(}{\lambda{(}{x}{)} {[}{\psi_{1}{(}{x}{)}}{]}}{)} {[}{A_{1}{(}{x}{)}\,^{\star}}{]}-{2}^{ - \frac{1}{2}} q_{1} \bar{\lambda{(}{x}{)}} \bar{\psi_{1}{(}{x}{)}} {[}{A_{1}{(}{x}{)}}{]}-\\{2}^{ - \frac{1}{2}} q_{2} {(}{\lambda{(}{x}{)} {[}{\psi_{2}{(}{x}{)}}{]}}{)} {[}{A_{2}{(}{x}{)}\,^{\star}}{]}-{2}^{ - \frac{1}{2}} q_{2} \bar{\lambda{(}{x}{)}} \bar{\psi_{2}{(}{x}{)}} {[}{A_{2}{(}{x}{)}}{]}+i \partial_{\mu}{(}{\bar{\psi_{1}{(}{x}{)}}}{)} \bar{\sigma}\,^{\mu} {[}{\psi_{1}{(}{x}{)}}{]} -\\ \frac{1}{2}q_{1} V{(}{x}{)}\,_{\mu} \bar{\psi_{1}{(}{x}{)}} \bar{\sigma}\,^{\mu} {[}{\psi_{1}{(}{x}{)}}{]}+i \partial_{\mu}{(}{\bar{\psi_{2}{(}{x}{)}}}{)} \bar{\sigma}\,^{\mu} {[}{\psi_{2}{(}{x}{)}}{]} - \frac{1}{2}q_{2} V{(}{x}{)}\,_{\mu} \bar{\psi_{2}{(}{x}{)}} \bar{\sigma}\,^{\mu} {[}{\psi_{2}{(}{x}{)}}{]}
    \end{array}
\end{equation}	

The above example can provide the reader with simple guidance on how to initiate programming in Cadabra specifically for supersymmetry use cases.	
Obviously, the real value of such automation or other types of computations will be much more noticeable when dealing with supersymmetric models in higher dimensions. For example, many fields emerge in supergravity theories in ten dimensions,  usually restricting us to checking supersymmetry only with respect to a subset of the transformation rules. With the help of Cadabra, heavy computations, especially in the context of field theory, can be addressed more easily.

\section*{Acknowledgments}
The author thanks Martin Cederwall and Daniel Butter for their helpful comments.


\begin{thebibliography}{99}

	

\bibitem{1}
K. Peeters, P. Vanhove, A. Westerberg, \emph{Supersymmetric higherderivative actions in ten and eleven dimensions, the associated superalgebras and
their formulation in superspace}, Class.Quant.Grav. 18 (2001) 843-890, [arXiv:hep-th/0010167].

\bibitem{2}
K. Peeters, A. Westerberg, \emph{The Ramond-Ramond sector of string theory
beyond leading order}, Class.Quant.Grav. 21 (2004) 1643-1666, [arXiv:hep-th/0307298].



\bibitem{3}
K. Peeters, \emph{Introducing Cadabra: a symbolic computer algebra system for field theory problems}, [arXiv:hep-th/0701238].


\bibitem{4}
K. Peeters, \emph{A field-theory motivated approach to symbolic computer algebra}, Comput. Phys.Commun. 176 (2007) 550 [arXiv:cs/0608005].

\bibitem{5}
L. Brewin, \emph{A brief introduction to Cadabra: a tool for tensor computations in general relativity}, Comput. Phys. Commun. 181 (3) (2010) 489–498 [arXiv:0903.2085].


\bibitem{6}
N. Ozdemir, M. Ozkan, O. Tunca, U. Zorba, \emph{Three-dimensional extended Newtonian (super) gravity}, JHEP 05 (2019) 130 [arXiv:1903.09377].

\bibitem{7}
K. Gubarev and E.T. Musaev, \emph{Polyvector deformations in eleven-dimensional supergravity}, Phys. Rev. D 103 (2021) 066021 [arXiv:2011.11424].


\bibitem{8}
D. Butter, J. Novak, M. Ozkan, Y. Pang and G. Tartaglino-Mazzucchelli, \emph{Curvature squared invariants in six-dimensional N = (1, 0) supergravity}, JHEP 04 (2019) 013 [arXiv:1808.00459].


\bibitem{9}
D. Butter, F. Ciceri and B. Sahoo, \emph{N = 4 conformal supergravity: the complete actions}, JHEP 01 (2020) 029 [arXiv:1910.11874].

\bibitem{10}
D. Butter, J. Novak and G. Tartaglino-Mazzucchelli, \emph{The component structure of conformal supergravity invariants in six dimensions}, JHEP 05 (2017) 133 [arXiv:1701.08163].



\end{thebibliography}
\end{document}